\documentstyle[twoside,epsf,epsfig]{article}
\input ibvs2.sty

\begin{document}
\IBVShead{5937}{29 April 2010}

\IBVStitle{A new ephemeris and an orbital solution of \boldmath{$\epsilon$}~Aurigae}

\IBVSauth{CHADIMA,~P.$^1$; HARMANEC,~P.$^1$; YANG,~S.$^2$; BENNETT,~P.D.$^3$; BO\v{Z}I\'C,~H.$^4$; RU\v{Z}DJAK,~D.$^4$; SUDAR,~D.$^4$; \v{S}KODA,~P.$^5$; \v{S}LECHTA,~M.$^5$; WOLF,~M.$^1$; LEHK\'Y,~M.$^6$; DUBOVSKY,~P.$^7$}

\IBVSinst{Astronomical Institute of the Charles University, Faculty of Mathematics and Physics, V~Hole\v sovi\v ck\'ach~2, CZ-180~00~Praha~8,
Czech Republic, e-mail: pavel.chadima@gmail.com}
\IBVSinst{Department of Physics and Astronomy, University of Victoria, P.O. Box 3055 STN CSC, Victoria, B.C., Canada V8W 3P6}
\IBVSinst{Department of Astronomy \& Physics, Saint Mary's University, Halifax, NS, B3H 3C3 Canada}
\IBVSinst{Hvar Observatory, Faculty of Geodesy, Ka\v{c}i\'ceva~26, 10000~Zagreb, Croatia}
\IBVSinst{Astronomical Institute of the Academy of Sciences, CZ-251~65~Ond\v{r}ejov, Czech Republic}
\IBVSinst{Severn\'i 765, CZ-500~03~Hradec~Kr\'alov\'e, Czech Republic}
\IBVSinst{Vihorlat Observatory, Mierova~4, Humenn\'e, Slovakia}

\SIMBADobjAlias{eps Aur}{HD 31964}
\IBVSabs{We collected rich series of RV measurements covering last 110 years}
\IBVSabs{and photometric observations from the past 6 primary eclipses,}
\IBVSabs{complemented them by our new observations and derived a new precise}
\IBVSabs{ephemeris and an orbital solution of epsilon Aur.}

\begintext
The bright star $\epsilon$~Aur (7 Aur, HD 31964, HR 1605; $V_{\rm max}$ = 3$^{\rm m}\!\!.$0; F0Ia+?) is an unusual eclipsing binary with a very long orbital period of 27.1
years (see Guinan \& Dewarf 2002 for a recent review). Its primary eclipse started in the summer 2009 and has naturally attracted the interest of many astronomers all over
the world. The aim of this paper is to present our analysis of an extensive collection of archival and new photometry, and radial velocities (RVs), and provide a new, more
precise ephemeris and orbital solution for the prediction of the current and future primary eclipses and a (not yet observed) secondary eclipse. Just prior to submission
of this paper, Stefanik et al.\ (2010; hereafter ST) published their analysis of a comparable dataset for this same star. ST presented a new orbital solution and improved
ephemeris for the binary but because the data analysis approach presented here is significantly different and may provide a more accurate ephemeris, we have proceeded to
publish our results also.

We compiled and digitized a large collection of RVs from the literature, including ST's dataset of 515 RVs obtained at the Harvard-Smithsonian Center for Astrophysics
(CfA). These data were augmented by our new series of electronic spectra from the Dominion Astrophysical Observatory (DAO) and the Ond\v{r}ejov Observatory. Altogether,
these RVs span an interval of 110 years. These RV observations are summarized in Table~1 and are plotted vs. time in Figure~1\footnote{RVs obtained during primary eclipse
were not used in our solution because they are known to deviate from purely orbital motion. These eclipse RVs are not included in Table~1 or Figure~1.}. We also collected
and digitized light curves from all six previously observed eclipses. Additionally, for the 2010 eclipse, we used standard photoelectric $V$ photometry obtained by PC,
HB, DR, DS and MW at Hvar Observatory, CCD $V$-band photometry obtained by ML at the Hradec Kr\'alov\'e Observatory, and visual observations by PD reduced to Johnson $V$
magnitude. These observations are listed in Table~2 and the individual eclipses are plotted in Figure~2\footnote{Some observations were not included because of their large
scatter and/or unsufficient coverage of a particular eclipse. We have also omitted extended datasets outside eclipse. These omitted data do not appear in either Table~2 or
Figure~2.}.

\begin{table}[ht]
\centerline{Table 1. Journal of available RVs.}
\begin{center}
\begin{tabular}{clrlccc}
\hline
from years & observatory & No. & reference\\
\hline
1899--1932 & Yerkes         & 298 & Frost et al. (1929) \\
1901--1913 & Postdam        & 173 & Ludendorff (1924) \\
1928--1958 & Mt.Wilson      & 53  & Struve et al. (1958)* \\
1970--1971 & Haute Provence & 18  & Castelli (1977) \\
1989--2009 & CfA            & 515 & Stefanik et al. (2010) \\
1994--2009 & DAO            & 99  & this paper** \\
2006--2009 & Ond\v{r}ejov   & 109 & this paper** \\
\hline
\end{tabular}
\end{center}
\begin{center}
{\footnotesize * RVs computed from the mean of 6 lines --- Fe~II~4123, Mg~II~4481, Fe~II~4508, Fe~II~4515, Fe~II~4576 and Fe~II~4629 \AA.}\\
{\footnotesize ** RVs computed from the mean of 5 lines --- Si~II~6347, Si~II~6371, Fe~II~6417, Fe~II~6433 and Fe~II~6456 \AA.}
\end{center}
\end{table}

\begin{table}[ht]
\centerline{Table 2. Journal of photometric observations during primary eclipses.}
\begin{center}
\begin{tabular}{clcrlccc}
\hline
mid-eclipse & observer & passband* & No. & reference\\
\hline
1848 & J.F.J.Schmidt            & pv      & 39  & Ludendorff (1912) \\
1875 & J.F.J.Schmidt            & pv      & 69  & Ludendorff (1912) \\
1902 & J.Plassmann              & pv      & 29  & Ludendorff (1903) \\
1902 & F.Schwab                 & pv      & 38  & Ludendorff (1903) \\
1929 & C.M.Huffer \& J.Stebbins & pe      & 98  & Huffer (1932) \\
1956 & K.Gyldenkerne            & $V$     & 131 & Gyldenkerne (1970) \\
1956 & G.Larsson-Leander        & $V$     & 106 & Larsson-Leander (1959) \\
1983 & J.L.Hopkins              & $V$     & 130 & Schmidtke (1985) \\
1983 & S.Ingvarsson             & $V$     & 119 & Schmidtke (1985) \\
2010 & Hvar Obs.                & $V$     & 100 & this paper \\
2010 & M.Lehk\'y                & $V$     & 21  & this paper \\
2010 & P.Dubovsk\'y             & pv($V$) & 28  & this paper \\
\hline
\end{tabular}
\end{center}
\begin{center}
{\footnotesize * Abbreviations `pv' and `pe' stand for photovisual and photoelectric, respectively.}
\end{center}
\end{table}

Here we provide more details of the new datasets. The DAO CCD spectra were obtained by SY and PDB and have a linear dispersion of 10~\AA~mm$^{-1}$ . The Ond\v{r}ejov CCD spectra
were obtained by PH, P\v{S}, M\v{S}, MW and a few other observers and have a dispersion of 17~\AA~mm$^{-1}$ . Both the DAO and Ond\v{r}ejov datasets cover the spectral region around
6300--6700~\AA. Their initial reductions were carried by SY and M\v{S} in IRAF\@. Rectification and RV measurements of the spectra were carried out by PC using the SPEFO
(Horn et al. 1996, \v{S}koda 1996) program's capability to compare direct and inverted line profiles. The zero point of the RV scale was determined  by measurement of
selected telluric lines (Horn et al. 1996). The Hvar dataset is actually $UBV$ photometry carefully reduced to the standard system (Harmanec, Horn and Juza 1994). The
Hradec Kr\'alov\'e CCD $BVRI$ photometry was obtained with a 2.8/29 Pentacon auto lens and SBIG ST-5C CCD camera. The visual estimates by PD, reduced to Johnson $V$-band
magnitude scale, were carried out using a modified version of Argelander's method developed by S.~Otero (\v{S}tefl et al. 2003). It is based on a cone vision and calibration
technique used to minimize the effects of extinction and colour differences. We are making all new RVs and photometric datasets available with the electronic version of
this paper\footnote{5937-t1.txt -- t5.txt}; the remaining RV and photometric data are already accessible from the electronic version of ST.

\IBVSedata{5937-t1}
\IBVSedata{5937-t2}
\IBVSedata{5937-t3}
\IBVSedata{5937-t4}
\IBVSedata{5937-t5}

\IBVSfig{4.3cm}{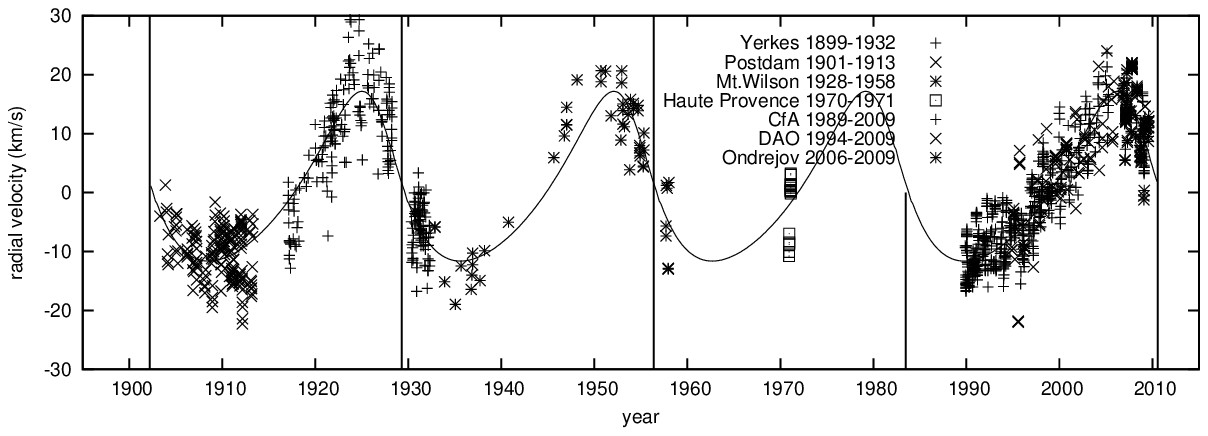}{RVs used in the orbital solution and the derived PHOEBE fit (curve). The vertical lines denote individual eclipses (during which RVs were not used in the solution). Plotted RVs are corrected for their individual $\gamma$ velocities.}
\IBVSfigKey{5937-f1.eps}{epsilon Aur}{radial velocity curve}

We used two independent programs, PHOEBE 0.31 (Pr\v{s}a \& Zwitter 2005) and FOTEL (Hadrava 2004), to derive new orbital solutions and formal light-curve solutions. All
data sets were assigned weights inversely proportional to the squares of their rms errors derived from preliminary solutions. In FOTEL, we allowed calculations of individual
systemic ($\gamma$) velocities for individual spectrographs. Since PHOEBE can treat only a single RV set, we used RVs with individual $\gamma$ velocities subtracted. It
turned out that the rms errors per observation for the RV sets in Table~1 were between 4 and 6~km\,s$^{-1}$. This indicates that the scatter is dominated by the intrinsic variations
of the F star because the actual measurement errors are typically less than 1~km\,s$^{-1}$. The RV solutions were used to derive the orbital eccentricity ($e$), longitude of periastron
($\omega$) and RV semiamplitude of a primary $K_1$, and the resulting values were then held fixed in the light-curve (LC) solutions. This is because the photometric data
used only covers orbital phases near primary eclipse and, therefore, these data do not constrain the eccentric orbit. LC solutions were used to derive an improved ephemeris,
assuming a mass ratio fixed at unity, and inclination fixed at $87^\circ$. The derived photometric period was held fixed again for the final iteration of the orbital solution,
evaluated using the {\sl unconstrained system} option in PHOEBE.

The final photometric ephemerides (based exclusively on the LC solutions) are:

\noindent $T_{\rm prim.min.}$ = HJD (2455402.8 $\pm$ 1.0) + (9890$^{\rm d}\!\!.$26 $\pm$ 0$^{\rm d}\!\!.$62) $\times$ $E$ (PHOEBE)\,,\\
$T_{\rm prim.min.}$ = HJD (2455403.7 $\pm$ 1.1) + (9890$^{\rm d}\!\!.$98 $\pm$ 0$^{\rm d}\!\!.$50) $\times$ $E$ (FOTEL)\,.

The epoch of primary minimum was allowed to vary independently for both the RV and LC solutions. We strongly prefer the more accurate value from photometry. For instance,
the epoch of the primary minimum from the final RV solution in FOTEL at HJD~2455347 differs significantly from the above ephemerides. ST arrived at the same conclusions from
their orbital solution; they obtained the epoch of the primary minimum at JD~2455136 (compared to their photometric minimum at JD~2455413). ST suggested that the gravitating
companion responsible for the orbital motion need not be the same as the extended gaseous structure responsible for the eclipses. However, they also noted that intrinsic
radial velocity variations in the F~supergiant's atmosphere might bias the orbital solution, thereby accounting for the discrepancy between the photometric and RV solutions.
We carried out an orbital solution in which the epoch of photometric mid-eclipse was held fixed and found that the resulting rms error was virtually identical to that of a
solution converged with the epoch free to vary. This result strongly suggests that the discrepancy is due to intrinsic RV variations of the F~supergiant and not due to asymmetry
in the companion's structure.

\IBVSfig{19cm}{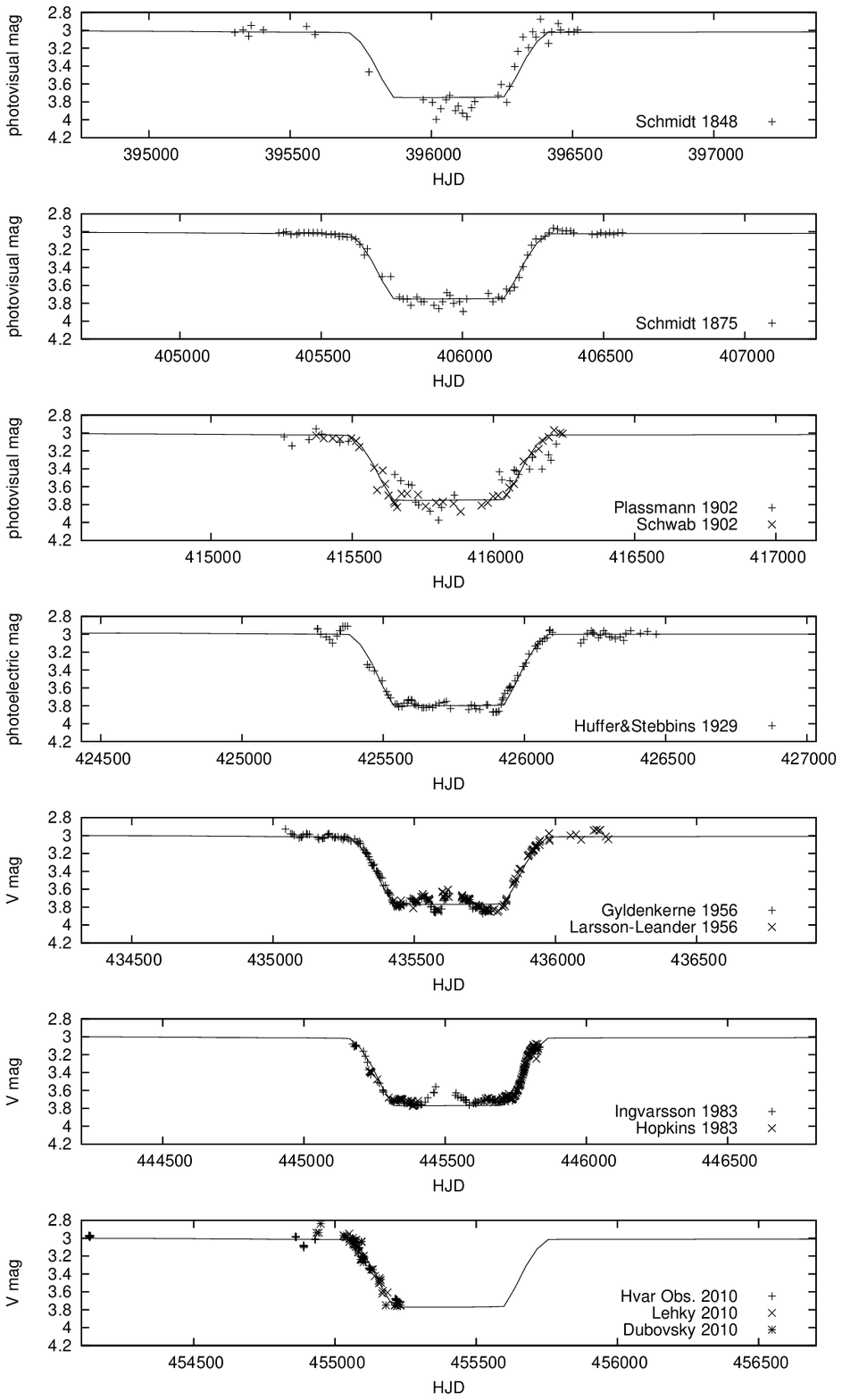}{Light curves from the last 6 eclipses, the current 2010 eclipse, and the PHOEBE fit (solid curve) are shown. Each measurement set is corrected to its individual 'zero level' magnitude. Mid-eclipse epochs have been centered using the new ephemeris and have been plotted on the same magnitude scale to facilitate  visual comparison.}
\IBVSfigKey{5937-f2.ps}{epsilon Aur}{light curve}

\begin{table}[ht]
\centerline{Table 3. New RV and LC solutions compared to those of Wright and Stefanik.}
\begin{center}
\begin{tabular}{lcccc}
\hline
element & PHOEBE & FOTEL & Wright & Stefanik\\
\hline
$T_{\rm periastron}$ & 2454596 $\pm$ 23*   & 2454622 $\pm$ 97*   & 2453130 $\pm$ 280*$\!\!_\#$ & 2454515 $\pm$ 80*$\!\!_\#$\\
$T_{\rm prim.min.}$  & 2455402.8 $\pm$ 1.0 & 2455403.7 $\pm$ 1.1 & 2455323$_\#$                & 2455413.8 $\pm$ 4.8\\
$T_{\rm sec.min.}$   & 2451681 $\pm$ 120*$\!\!_+$  & 2451610 $\pm$ 180*$\!\!_+$ & --           & --\\
$P$ (d)              & 9890.26 $\pm$ 0.62  & 9890.98 $\pm$ 0.50  & 9890 (assumed)              & 9896.0 $\pm$ 1.6\\
$e$                  & 0.256 $\pm$ 0.012   & 0.249 $\pm$ 0.015   & 0.200 $\pm$ 0.034           & 0.227 $\pm$ 0.011\\
$\omega$ ($^\circ$)  & 41.2 $\pm$ 3.1      & 43.3 $\pm$ 4.0      & 346$\pm$11                  & 39.2 $\pm$ 3.4\\
$K_1$ (km\,s$^{-1}$) & 14.40 $\pm$ 0.38    & 14.30 $\pm$ 0.25    & 15.00 $\pm$ 0.58            & 13.84 $\pm$ 0.23\\
\hline
\end{tabular}
\end{center}
\begin{center}
{\footnotesize * Errors from RV solutions. $^+$ Errors are semianalytical estimates. $^\#$ Epochs recalculated for the authors' original periods.}
\end{center}
\end{table}

\vskip -5mm

\IBVS2fig{5cm}{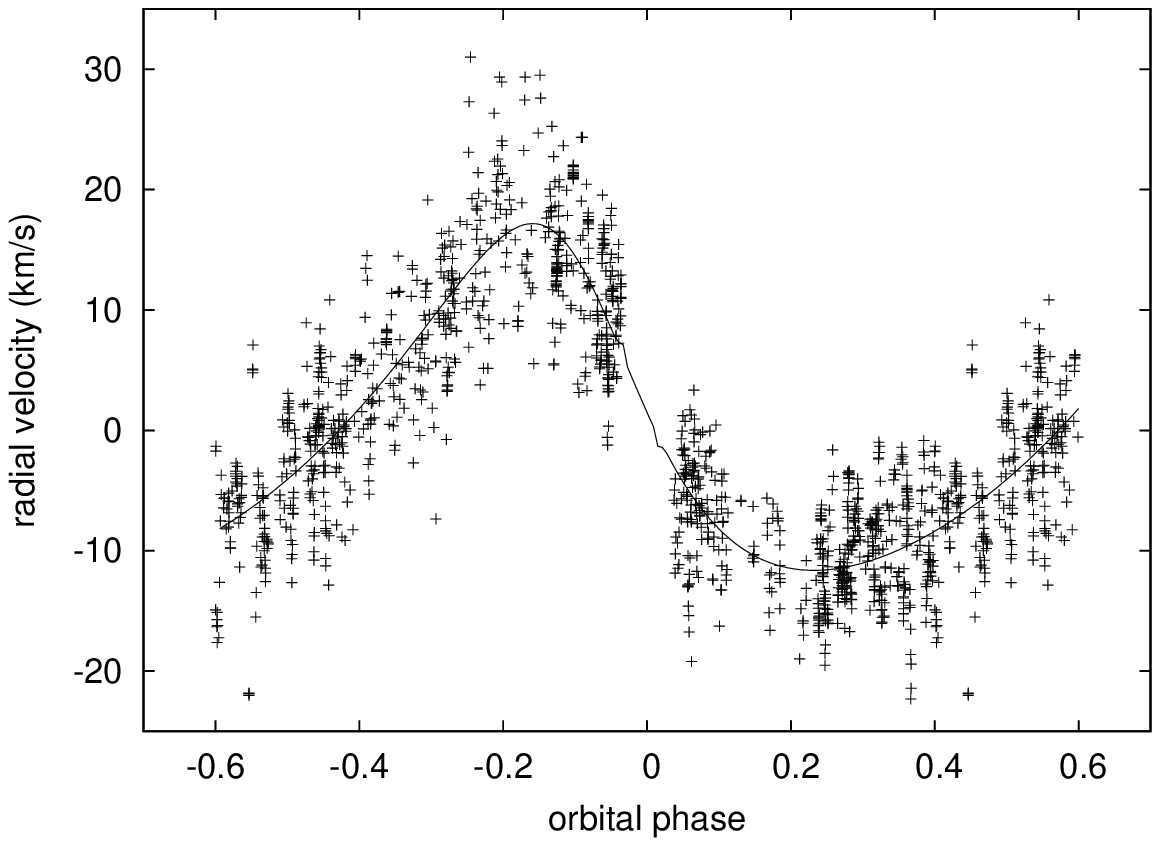}{A phase plot of all RVs used in the orbital solution, and the derived PHOEBE fit (solid curve). Plotted RVs have been corrected for their individual $\gamma$ velocities. Note that PHOEBE also accounts for rotational effect during eclipse.}{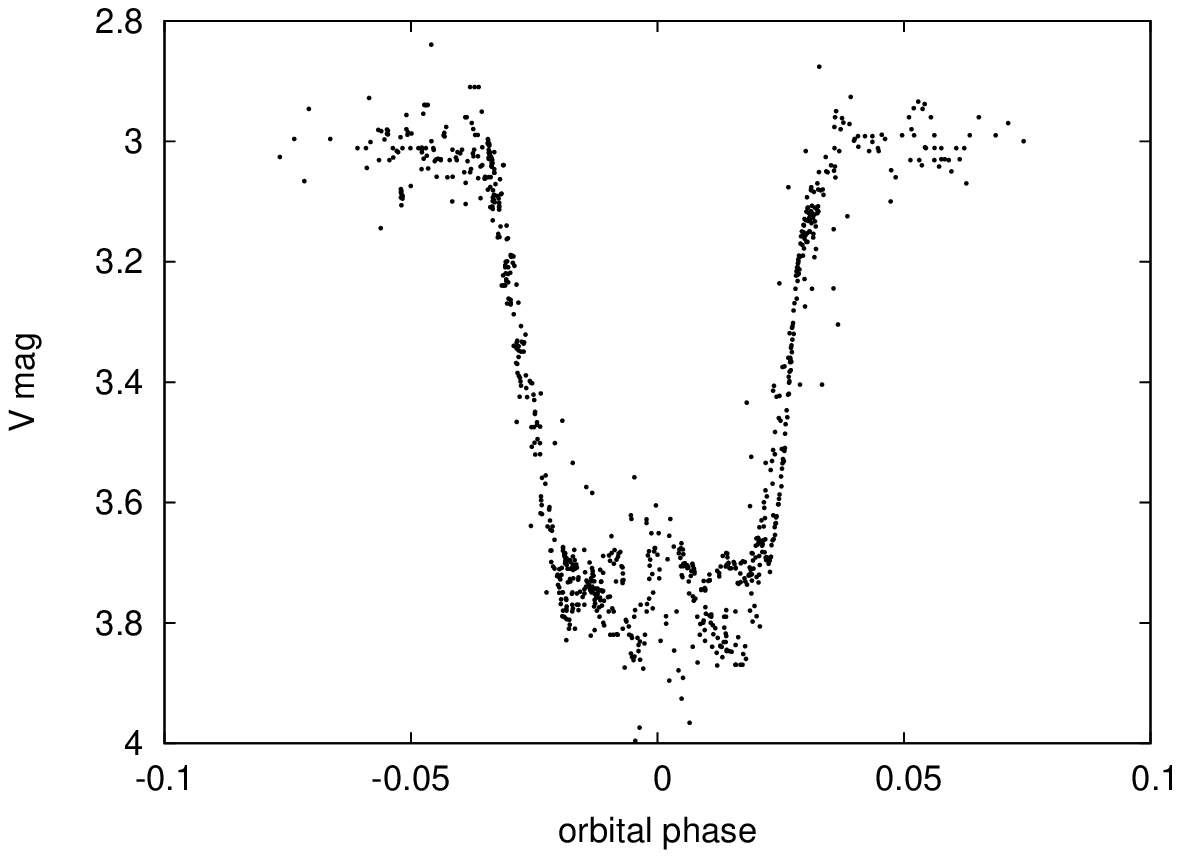}{A phase plot of all photometric observations used in the ephemeris calculation. Each measurement set has been corrected for individual 'zero level' magnitude.}
\IBVSfigKey{5937-f3.eps}{epsilon Aur}{phased radial velocity curve}
\IBVSfigKey{5937-f4.eps}{epsilon Aur}{phased light curve}

We present our orbital solutions in Table~3, along with those of Wright (1970) and ST\@. A phase plot, using our new ephemeris from PHOEBE, of all RVs and photometry is shown
in Figure~3. Note that our new solutions, obtained with two independent programs, agree within their respective errors. Our ephemerides predict the next primary mid-eclipse
will occur on July 25-26, 2010, and the next secondary mid-eclipse in 2027.  The previous secondary eclipse should have occurred in 2000. When compared to Wright, we obtain
a significantly different orientation of the orbit in space (longitude of periastron $\omega$), a higher eccentricity ($e$), and a different epoch of the primary minimum.
Our results are much closer to the ST solution, but we still disagree with ST by more than the estimated errors. At the request of the referee, we mention that the resulting
relative photometric radii from the LC solutions were 0.045 and 0.216 from PHOEBE, and 0.058 and 0.218 from FOTEL\@. We caution the reader, however, not to give these values
much weight since neither program can treat disks; both assume two stellar bodies.

Two important conclusions about $\epsilon$~Aur, which disagree with the generally accepted model, follow from our study:

\smallskip
1. Inspection of Figs. 2 and 3 shows that the idea of a central brightening inside the eclipse, interpreted as evidence of a hole in the disk (see, e.g., Carroll et al. 1991),
should be reconsidered. Note that the `flat' part of each recorded eclipse is different and what is seen are most probably the physical light variations, similar to the
out-of-eclipse variability. Of course, the final conclusion will come from a detailed analysis of colour changes and other types of observations and from the photometry
secured this summer.

\smallskip
2. The right panel of Figure~3 shows that claims of variability in the width and duration of individual observed eclipses, which have been used to infer a decline in the
primary's radius over time (Saito 1986), are not supported by the data. It is apparent that the cyclic but irregular physical light variations affected the different eclipses
differently. It will be difficult to obtain a `pure' eclipsing light curve without a better understanding and quantitative description of these light changes.

\medskip
{\it Acknowledgements.} We acknowledge the use of the programs PHOEBE and FOTEL made available by their authors Drs.~Andrej~Pr\v{s}a and Petr~Hadrava. We profited from the use
of the bibliography maintained by the NASA/ADS system and the CDS in Strasbourg. We would like to express our admire and gratitude to our predecessors who carefully accumulated
a large body of observational data used in this study. Our special thanks go to Mr.~Jeff~Hopkins for his observations and the creation of the web page with detailed and updated
information on $\epsilon$~Aur (www.hposoft.com/Campaign09.html). Drs.~A.~Kawka and P.~Mayer and students E.~Arazimov\'a, B.~Ku\v{c}erov\'a and J.~Polster obtained a few
Ond\v{r}ejov spectra. ML acknowledges the use of a telescope with a CCD camera of the Hradec Kr\'alov\'e Observatory and Astronomical Society of Hradec Kr\'alov\'e and the
help of Dr.~M.~Bro\v{z} with the reductions. Our thanks are also due to Dr.~Martin~\v{S}olc for his help with a translation of papers written in German. The Czech authors
were supported by the grants 205/06/0304, 205/08/H005, and P209/10/0715 of the Czech Science Foundation and also from the research programs AV0Z10030501 and MSM0021620860.

\references

Carroll, S.M. et al., 1991, {\it ApJ}, {\bf 367}, 278. 

Castelli, F., 1977, {\it ApSS}, {\bf 49}, 179. 

Frost, E.B. et al., 1929, {\it Publ. Yerkes Obs.}, {\bf 7}, 81. 

Guinan, E.F., Dewarf, L.E., 2002, {\it ASP Conf. Series}, {\bf 279}, 121. 

Gyldenkerne, K., 1970, {\it Vistas in Astronomy}, {\bf 12}, 199. 

Hadrava, P., 2004, {\it Publications Astron. Inst. Czechosl. Acad. Sci.}, {\bf 92}, 1. 

Harmanec, P., Horn, J., Juza, K., 1994, {\it A\&AS}, {\bf 104}, 121. 

Horn, J. et al., 1996, {\it A\&A}, {\bf 309}, 521. 

Huffer, C.M., 1932, {\it ApJ}, {\bf 76}, 1. 

Larsson-Leander, G., 1959, {\it Arkiv Astron.}, {\bf 2}, 283. 

Ludendorff, H., 1903, {\it AN}, {\bf 164}, 81. 

Ludendorff, H., 1912, {\it AN}, {\bf 192}, 389. 

Ludendorff, H., 1924, {\it Sitzungsber. der Preuss. Akad. der Wissensch.}, 49

Pr\v{s}a, A., Zwitter, T., 2005, {\it ApJ}, {\bf 628}, 426. 

Saito, M., Kitamura, M., 1986, {\it Astrophys. and Space Science}, {\bf 122}, 387. 

Schmidtke, P.C., 1985, {\it eepa.rept}, 67. http://www.hposoft.com/Astro/PEP/EAur82-88.html \BIBCODE{1985eepa.rept...67S}

\v{S}koda, P., 1996, {\it ASP Conf. Series}, {\bf 101}, 187. 

Stefanik, R.P. et al., 2010, {\it AJ}, {\bf 139}, 1254. 

\v{S}tefl, S. et al., 2003, {\it A\&A}, {\bf 402}, 253. 

Struve, O. et al., 1958, {\it ApJ}, {\bf 128}, 287. 

Wright, K.O., 1970, {\it Vistas in Astronomy}, {\bf 12}, 147. 

\endreferences

\end{document}